\documentclass{JINST}

\usepackage{natbib}

\usepackage{graphics}
\usepackage{graphicx}   
\usepackage{subfigure}
\usepackage{amssymb}
\usepackage{amsmath}
\usepackage{booktabs}

\title{The High Voltage Feedthroughs for the ATLAS Liquid Argon Calorimeters}

\author{B. Botchev, G. Finocchiaro, J. Hoffman, R.L. McCarthy, M. Rijssenbeek\thanks{Corresponding author}, J. Steffens, A. Talalaevskii\thanks{present address: HYPRES Inc., NY}, 
M. Thioye, M. Zdrazil\thanks{present address: Lawrence Berkeley National Laboratory, CA}\\Stony Brook University, Stony Brook, NY 11794-3800\\
\email{michael.rijssenbeek@stonybrook.edu}}
\author{J. Farrell, S. Kane\\Brookhaven National Laboratory, P.O. Box 5000
Upton, NY 11973-5000}

\abstract{
The purpose, design specifications, construction techniques, and testing methods are described for the high voltage feedthrough ports and filters of the ATLAS Liquid Argon calorimeters. These feedthroughs carry about 5000 high voltage wires from a room-temperature environment (300~K) through the cryostat walls to the calorimeters cells (89~K) while maintaining the electrical and cryogenic integrity of the system. The feedthrough wiring and filters operate at a maximum high voltage of 2.5~kV without danger of degradation by corona discharges or radiation at the Large Hadron Collider.%
}


\begin{document}
\bibliographystyle{plain}

\section{Introduction}
The Large Hadron Collider (LHC), located at CERN near Geneva, Switzerland, is a proton-proton collider with a 14~TeV center of mass energy and a design luminosity of $10^{34}~\rm {cm^{-2}s^{-1}}$. The extreme operating conditions of the LHC impose severe constraints on detectors, in terms of radiation tolerance, background rejection capability, response speed, spatial coverage, time stability and reliability. 
 
The ATLAS (A Toroidal LHC ApparatuS) experiment\cite{AtlasProposal} presently under construction, is one of five experiments at the LHC which will start operation in 2008. This multi-purpose detector with a wide physics program was specifically designed and optimized for the discovery of new particles predicted by the Standard Model, such as the Higgs boson, and new physics such as supersymmetric particles, but also to improve precision measurements of gauge bosons, such as the $W^{\pm}$ bosons, and heavy quark properties. ATLAS has the usual ``onion-layer'' structure composed of  independent detector subsystems with the innermost being the inner detectors (Silicon tracker, Pixels and Transition Radiation Tracker), followed by the calorimeters (Electromagnetic (EM) and Hadronic), and finally the muon detectors. The calorimeters will play a key role in measuring energy, position, and time of electrons, photons, and jets.

The ATLAS EM calorimeter is a lead-liquid argon (LAr) sampling detector with an accordion geometry\cite{RD3}\cite{LArTDR} which  guarantees full azimuthal acceptance as well as a coverage in the pseudorapidity region of $|\eta| < 3.2$. It is divided into one barrel ($|\eta|<1.475$) and two end-caps ($1.375<|\eta|<3.2$) and is segmented in depth in 4 compartments. Those are, respectively, the presampler, the strips, the middle, and the back samplings. Because the EM and Hadronic calorimeters have to match challenging requirements for energy, position, and time resolutions, LAr was chosen for its intrinsic linear behavior, response stability, and radiation tolerance. 

\section{Conceptual design of the high-voltage feedthrough port}
Feeding high-voltage (HV) wires into the LAr calorimeters presents different problems from those which must be solved in order to get the signals out of the LAr (at 89 K). The HV lines are especially susceptible to problems resulting from condensation near the exposed surfaces, so it is highly desirable to maintain the wire feedthroughs at ambient temperature (and if necessary heat them) in order to avoid condensation. The total cross section allotted to HV supply wires is severely limited as most of the available space around the edges of the cryostats is taken up by signal feedthroughs and cryogenic services. There is no requirement that the HV lines should enter the cryostat with axial symmetry (like the signal wires). Thus, the HV feedthroughs (HVFTs) are placed at the highest point on the cryostat and are kept at room-temperature, with a buffer volume of gaseous argon gas directly below the wire feedthroughs separating them from the liquid argon below. Being non-cryogenic simplifies the design of a very dense HV feedthrough greatly. The heat flow through the HVFT is limited by the choice of constantan conductor and by thin-walled metal bellows that flexibly link the outer and inner cryostat vessels. 

\begin {figure} 
\centering
\includegraphics[height=5.0 in]{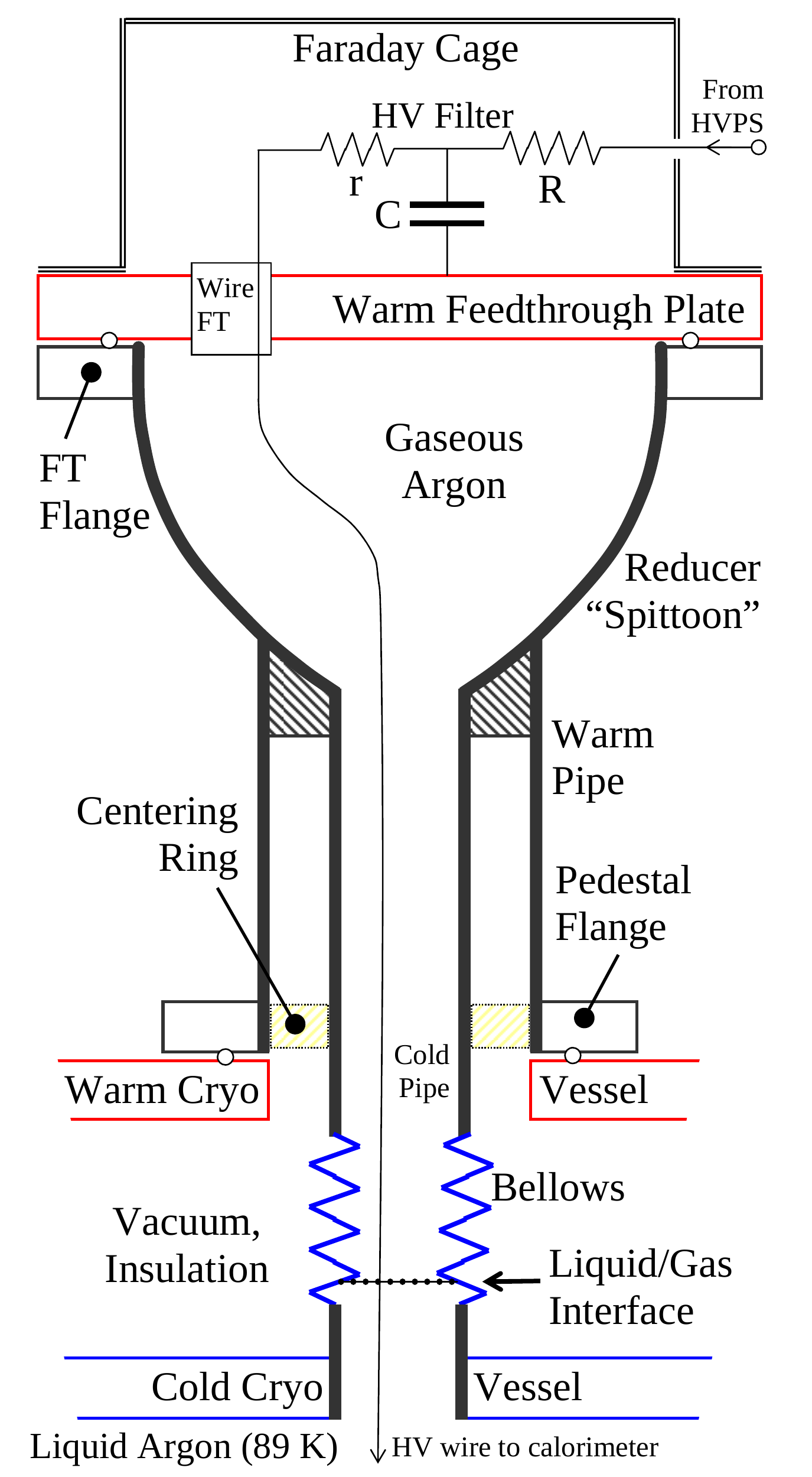}
\caption{Conceptual drawing of the ATLAS LAr high voltage feedthrough (HVFT) port and the HV filters. The HVFT ports are located near the top of the LAr cryostats. 
Not to scale.\label{fig:HVFTconcept}}
\end{figure}

The conceptual design of a HVFT port is shown in Fig.~\ref{fig:HVFTconcept}. The HV lines lead from the various calorimeter patch panels and come up through the surface of the LAr into a small pipe (73~mm inner diameter). A stainless steel bellows in this cold pipe accommodates the differential motion of the cold and warm walls during cooldown and serves as the dominant thermal resistance in the cold pipe. At approximately 0.9~m above the liquid, the cold pipe is welded to the warm pipe of the HVFT which is bolted to the outer cryostat vessel so that, above this point, the FT wall is at room temperature. Above this point, the HV wires fan out to four HV Wire FTs (WFTs) mounted in a strong FT Plate, from where each wire is connected to its own RC filter to reduce electronic noise entering the cryostat. The HV filter section is also severely limited by the available space. A total of 840 HV channels is accommodated in each HVFT port. Two HVFT ports serve the barrel cryostat, one per end, and two ports serve each of the two end-cap cryostats. This makes for a total of six HVFT ports with 5040 HV channels, including about $10\%$ spare channels.

Each HVFT port was constructed and installed in three parts: (1) the mechanical/cryogenic part consisting of the port proper; (2) the electrical part consisting of the FT plate with the four WFTs with HV wire bundle and HV backplane; and (3) the HV filter crate with filtermodules. The ports were welded, installed, and first tested as part of the cryostat commissioning. The FT plate with HV wiring was installed later, at the time the calorimeter modules were installed in the cryostats. Finally, the HV filters were installed last after the cryostats were closed and HV testing in LAr was started. 

\begin {figure}  
\centering
\includegraphics[height=3.6 in]{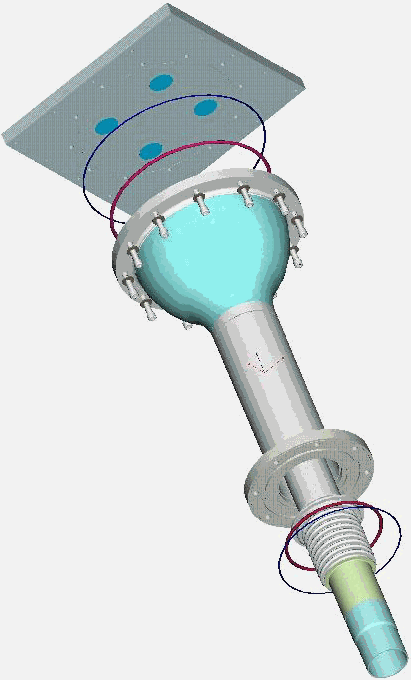}
\hfil
\includegraphics[height=3.6 in]{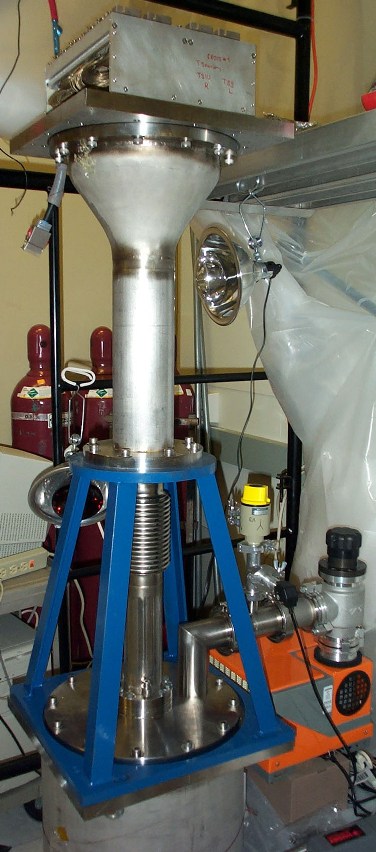}
\caption{Mechanical design of a HV feedthrough port (left). The four holes on the FT plate accommodate the four wire feedthroughs of 210 wires each. Also shown are the O-rings and RF gaskets. The leak testing at Stony Brook of a completed HVFT port mounted on a test vessel which contains the HV wire bundle (right).\label{fig:HVFTport}}
\end{figure}

\section{Mechanical Construction of the HVFT ports }

The mechanical design and technical specifications of one HVFT port are respectively shown in Fig.~\ref{fig:HVFTport} and Table~\ref{tab:HVFTdimensions}. A single port consists of several mechanical parts that are welded or bolted together and meet stringent cryogenic code requirements.

\begin{table}[t!]
\begin{center}
\begin{tabular}{||l|c|c|c|c|c|}
\hline
Component & Height & Thickness & Inner ${\o}$ & Outer ${\o}$ & Stain. steel type \\  
\hline
Bellows & 140 & 0.3 x 2 & 80.0 & 101.0 & 331 \\ 
\hline
Cold pipe & 424 (ab) & 1.6 & 73.0 & 76.2 & 304L \\
\hline 
Warm pipe & 333.5 & 2.11 & 110.1 & 114.3 & 304L/316L \\ 
\hline
	        &         & 2.1 & 110.1 & 114.3 &  \\ 
Reducer &  178 &  to  &    to     &     to    & 304L\\	
                 &         & 3.4 & 266.3 & 271.3 & \\
\hline
Plate &  & 25 &  &  & 304L \\ 
\hline
\end{tabular}
\end{center}
\caption{Dimensions (mm) of HVFT port components (ab means above bellows).\label{tab:HVFTdimensions}}
\end{table}

\begin {figure} 
\centering
\includegraphics[width=5.5 in]{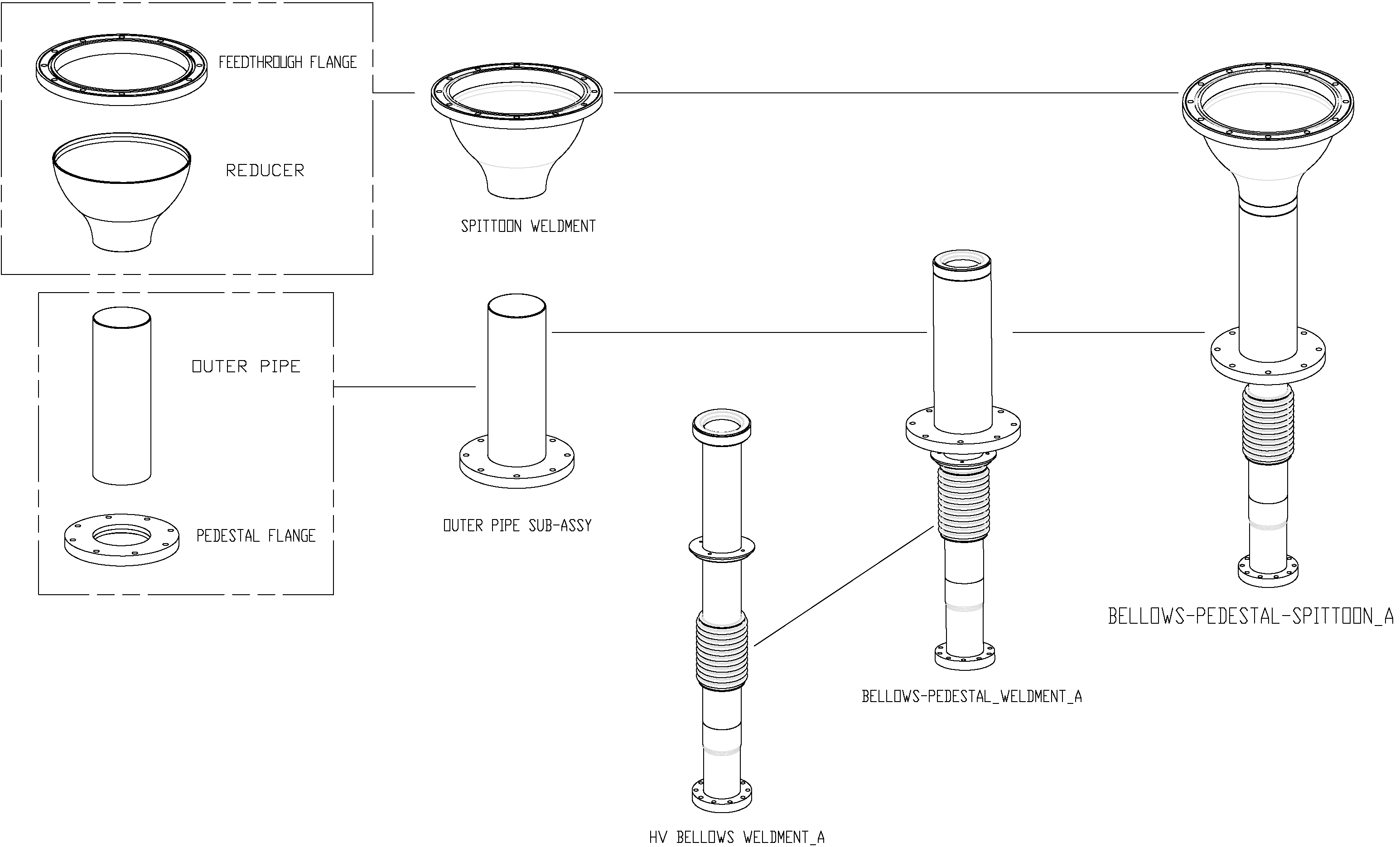}\\ 
\includegraphics[width=5.5 in]{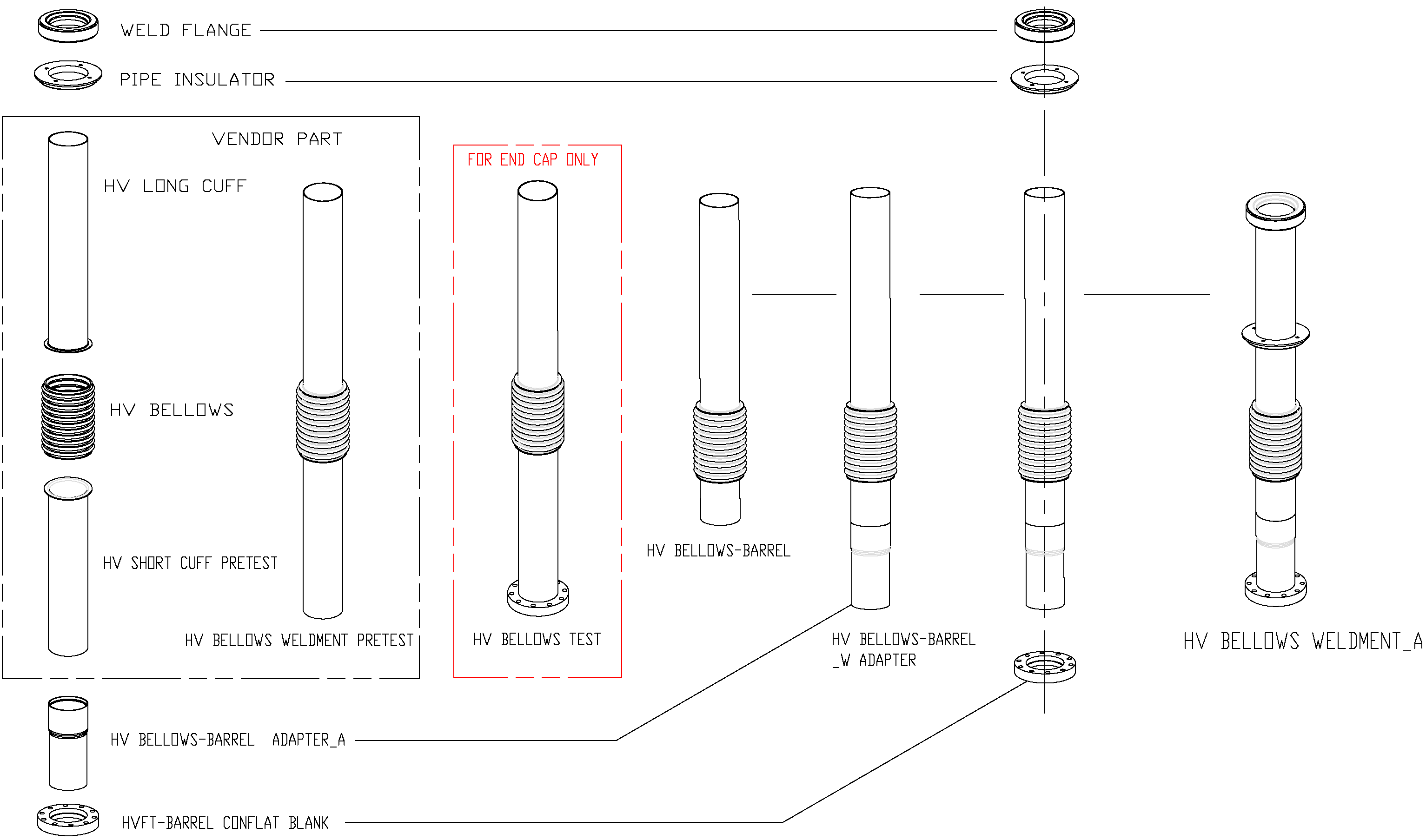}
\caption{Assembly tree  of a HVFT port: cold pipe and bellows (below). Assembly of the warm parts: pedestal, outer pipe, reducer; and the final port assembly (above).\label{fig:AssemblySeq}}
\end{figure}

Depending on its location, end-cap or barrel, the bellows is first welded on either side to a long and a short cuff (end-cap) or to a long cuff and a barrel adapter (barrel), see Fig.~\ref{fig:AssemblySeq}. A PEEK (Poly-Ether-Ether-Ketone) pipe insulator is slid on the long cuff (which provides centering support of the cold pipe within the warm pipe), with a weld flange capping the top end of the cold pipe and bellows assembly. On the bottom a temporary test flange is welded. This end is cut to size at the time of installation, when the HVFT port is welded to an aluminum-to-stainless steel transition piece on the cold cryostat.

The bellows has a two-ply structure and consists of two individual bellows, one inside the other. The thickness of each ply is 0.3 mm. The bellows are guaranteed to survive more than 1000 cooldown cycles, and can operate in LAr and in vacuum. 

The warm components (the transition flange, the reducer, the warm pipe, and the pedestal flange) are welded together with the cold assembly to form the completed HVFT unit. The various pieces were all machined at Stony Brook University, while all welding was executed by a code-certified welder at BNL.

\begin {figure}  
\centering
\includegraphics[width=3.0 in]{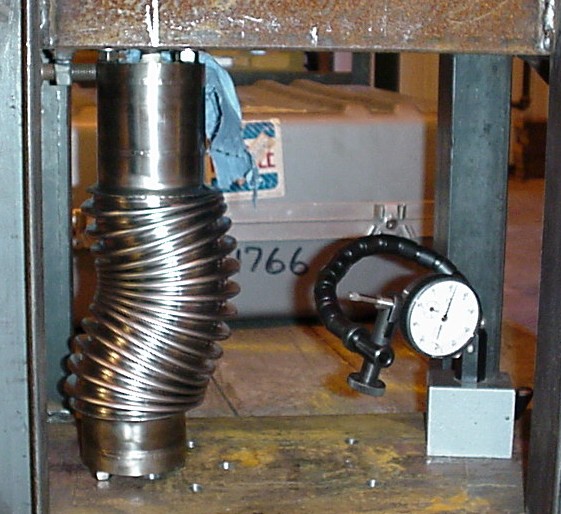}
\caption{``Squirm'' test of a bellows unit. Squirm occured at 190 psi (13 atm).\label{fig:SquirmTest}}
\end{figure}
 
In addition to the weld joints, there are several O-ring joints in each HVFT port structure: at the SS pedestal flange between the HVFT port and the warm aluminum cryostat vessel, and at the joint between the FT transition flange and the FT plate. Both these flanges contain an O-ring and a RF gasket. Since electrical noise considerations require that the HVFT port be an integral part of the Faraday cage with the cryostat, a CuBe wire mesh RF gasket (Laird Technologies part no.~8300-0225-40) is used to ensure a good electrical connection. The O-ring material is EPDM rubber (Ethylene Propylene Diene Monomer). Finally, each 210-wire FT is sealed in the FT plate with a nitrile O-ring made of BUNA-N (Acrylonitrile Butadiene Terpolymer). 

\subsection{Performance and Tests}

In the HVFT port there are two dominant sources of heat leaks into the LAr: the cold pipe and the constantan-PEEK HV wire bundle. The heat flow along the cold wall is mostly determined by the stainless steel bellows (having the smallest wall thickness and therefore the smallest heat conduction) and amounts to 0.8~W. The LAr is expected to rise inside the bellows to a level a few centimeters above the stainless steel - aluminum transition to the cold cryostat wall. The heat flow along the HV wire bundle is 0.4~W. So the total heat flow is about 1.2~W, making it unnecessary to heat the FT plate unless a major gas leak occurs in one of the HVFT seals causing the LAr level to rise. Thus, in order to control the temperature and heat the FT plate if necessary, two thermocouple pairs and four resistive heaters are mounted on the FT plate. Temperature readout and heater power wires are filtered and lead to the ATLAS cryogenics control room. 

Several tests were performed to check the weldments and the O-ring seals. 
A FT plate outfitted with four 210-wire FTs and bolted to the reducer was tested and found to withstand an absolute pressure $P_{abs}\geq 68$ bar without leakage and deformation. 

Given the highly sensitive role they have in the HVFT mechanics, the bellows were tested separately at BNL to test the ``squirm''-limit (sideways bulging instability that could destroy the bellows), see Fig.~\ref{fig:SquirmTest}. The critical pressure for squirm, $P_{\rm cr}$, was found as $P_{\rm cr}>4\times P_{\rm op}$, where $P_{\rm op}$ is the nominal operating pressure of 2.8~bar.

In order to ensure the vacuum integrity of the high voltage FTs, every FT port, each containing four wire FTs, was required to have a leak rate below $10^{-9}$ STD cc/s of Helium at standard temperature and pressure. Measurements were conducted at Stony Brook. The expected permeation rate L in STD cc/s through the seal is approximated by the following formula\footnote{NASA, {\tt www.nasa.gov/offices/oce/llis/0674.html}} in metric units: 
\[L \approx 0.123 \times \mathcal{P} \times ID \times \bigtriangleup P \times Q \times (1-S)^{2}\]
where $\mathcal{P}$ is the permeation rate (STD $\rm cc/s \times cm/cm^{2}bar$) of the gas through the seal at the anticipated temperature, ID (in cm) the inner diameter of the O-ring, $\bigtriangleup P$ (in bar) is the pressure difference across the seal, Q the quality factor of the seal (about 1.25 for a non-lubricated O-ring at $30\%$ squeeze), and S is the squeeze factor expressed as a decimal. We calculated a helium permeation rate $L_{He} = (4-6)\times10^{-6}$ STD cc/s for a nitrile seal, and $L_{He} = 15\times10^{-6}$ STD cc/s for an EPDM seal.

\begin{table}[htbp]
\centering
  \begin{tabular}{|l|c|c|c|c|}
    \hline
    Gas & Nitrile(BUNA-N) & Butyl & EPDM & Butadiene \\ 
    \hline
    He & (5.2-8)$\times10^{-8}$ & (5.2-8)$\times10^{-8}$ & 20$\times10^{-8}$ & 12$\times10^{-8}$ \\ 
    \hline
    A & (1.6-3.9)$\times10^{-8}$  & 1.2$\times10^{-8}$  & (11-23)$\times10^{-8}$  & NA \\
    \hline 
    O$_2$ & (0.7-6)$\times10^{-8}$  & (1.0-1.3)$\times10^{-8}$  & NA & (8-14)$\times10^{-8}$  \\ 
    \hline
    H$_2$O & 760$\times10^{-8}$  & (30-150)$\times10^{-8}$  & NA & NA \\ 
    \hline
  \end{tabular}
  \caption{Permeation [STD $\rm cc/s \times cm/cm^{2}/bar$] of O-ring material at $20-30^{o}$C.
  \label{tab:PermeationRates}}
\end{table}

Here, we show measurements involving both types of O-rings that were made in a 10-day interval. In the tests, a HVFT port was continuously pumped out for two days to bring the pressure below 4~mTorr. An oil-free membrane roughing pump to back up a high vacuum molecular turbo pump was used. The main pump was then shut off and the helium leak detector (incorporating its own diffusion pump) switched onto the system. Helium background, as measured with the leak detector, stabilized in $\sim1$~hr at about ($1.4\pm0.2)\times10^{-9}$ STD cc/s. 

We measured individual leak rates of both types of O-ring seals by enclosing them in bags that were filled with Helium. No Helium was measured above the background $1.3\times10^{-9}$ STD cc/s for 5 minutes for the wire FTs which contained the small nitrile O-rings. Next, helium was injected in a second bag enveloping the large EPDM seal between the FT plate and FT flange. Starting after 1-2 minutes, a slow and steadily accelerating rise of helium leakage at a rate of ($0.5-1)\times10^{-9}$ STD cc/s per minute was observed. After 15 minutes of flow around the EPDM seal, the helium supply was shut off and the bag removed. About 5 minutes later the measured leak rate was still increasing and did not reach a plateau during the observation period. At the end of observations, the Helium leak rate was $\sim6\times10^{-8}$ STD cc/s and still rising at $1\times10^{-8}$ STD cc/s per minute. 

The delayed onset of Helium detection is in agreement with slow permeation through the large EPDM seal in the FT plate (see Table~\ref{tab:PermeationRates}). The observed background-subtracted permeation of this seal is $1\times10^{-9}$ STD cc/s after 5 minutes. The  smaller seals of the WFTs did not show signs of permeation (less than $2\times10^{-10}$ STD cc/s for about 5 minutes); for these seals we expect a permeation rate one order of magnitude smaller (a factor 2 less because of size, and probably a factor 4 less because of the lower permeability of Buna-N compared to EPDM). 

\begin {figure} 
\centering	
\includegraphics[height=3.0 in]{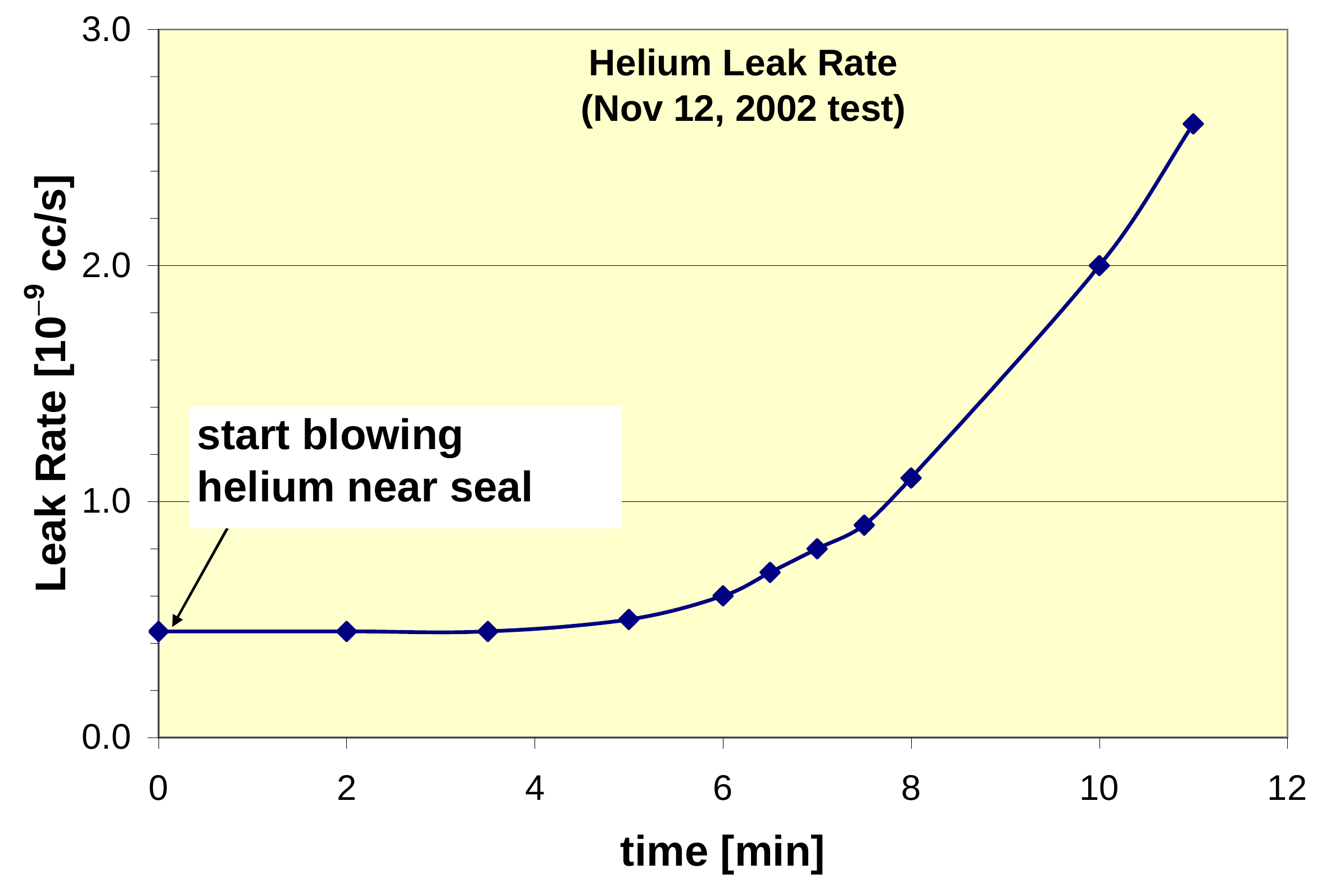}
\caption{Helium leak rate in a HVFT port (results of second test).\label{fig:HeLeakTest}}
\end{figure}

A second type of test, performed on all HVFT ports, involved ``spraying'' helium only near the various seals as is customarily done for UHV seal testing. Helium was sprayed (from a helium line ending in a hollow needle) at a rate of $\sim10$ mL/s ($\sim5$ bubbles/s as measured by injecting it in a fluid). The needle was passed along the seals at a speed of 
$\sim2$~cm/s. 
For this test, the helium leak detector was connected at the output of the turbo-molecular pump and also served as roughing pump. Measurements started after a baseline/background plateau of $0.45\times10^{-9}$ STD cc/s was reached after $\sim0.5$ hrs of pumping. No increase of the helium leak rate was observed during the measurement, which lasted several minutes. In an attempt to confirm the permeation effect observed in the earlier test, helium was finally injected at high flow rate in a bag enveloping all seals. A steady increase was observed, as in the first test, starting after 4-5 minutes, see Fig.~\ref{fig:HeLeakTest}.

Although helium permeation through the HVFT seals is significant, no measurable leaks were detected in any of the FT plates. The expected $O_{2}$ permeation rate, assuming $P = 5\times10^{-8}$ and $\bigtriangleup P = 0.2$ atm, equals $LO_{2} \leq 8.8\times10^{-7}$ STD cc/s = $8.8\times10^{-7} /22413$ cc/mole = $3.9\times10^{-11}$ mole of $O_{2}$/s = $1.2\times10^{-3}$ mole of $O_{2}$/yr, again for the Nitrile seal. For a total LAr load of the half-barrel of 30 tonnes = $3\times107$ g/40 g/mole = $7.5\times105$ mole of Ar, this represents a pollution of $\sim1.6$ ppb or less of $O_{2}$ per year. Oxygen diffusion into the LAr (at an inside pressure of $\sim3.5$ atm) will be several orders of magnitude smaller  than the helium permeation rate (with vacuum inside the HVFT), and is thus not a concern. Moreover, because the pressure is reversed in ATLAS, the rate of diffusion (not the rate of permeation) of oxygen across the seals is a more accurate measurement of pollution and is several orders of magnitude lower.
 
\subsection{The HV wire}

A single HV wire is used to carry each high voltage channel through the FT plate. The characteristics of this wire are shown in Table~\ref{tab:HVwire}. The HV wire has a solid constantan conductor and an extruded PEEK insulator. We chose wire with a constantan conductor because of its relatively low thermal conductivity, thus reducing the heat leak along the wires. Constantan was favored over stainless steel (an even poorer thermal conductor) because it is relatively easy to solder. PEEK insulation was chosen both because of its very high radiation tolerance and its very high electrical resistance. PEEK can be used throughout the calorimeters, even in the forward region which receives the highest radiation dose. Constantan wire with Kapton insulation was also considered but rejected because of its lower radiation tolerance and because of the possibility that, in principle, its insulation could unravel. All wires used in the construction of the FTs were purchased from HABIA Cable\footnote{HABIA Cable, {\tt www.habia.se/}}.

\begin{table}[htbp]
  \centering 
  \begin{tabular}{|l|c|c|}
    \hline 
    Characteristics & Material/Specifications \\ 
    \hline
    Conductor 0.41~mm ${\o}$ (26 AWG) & constantan (60\% Cu, 40\% Ni) \\ 
    Insulator 1.01~mm ${\o}$ & PEEK (Poly-Ether-Ether-Ketone) \\ 
   \hline	
      &   \\ 
   \hline
    Minimum breakdown voltage & 5.0 kVAC (tested to 15~kVAC at HABIA)  \\ 
    Radiation tolerance & $10^{7}$ Gy ($10^{9}$~rad)\\ 
    Thermal conductivity (constantan) & 17~W/m.K at 77~K (22~W/m.K at 273~K) \\ 
    Electrical resistance (constantan) & 3.82 $\rm \Omega/m$ \\ 
    Electrical resistivity (PEEK) & $\rho_{R}\geq1.7$ $\times10^{13} \rm \Omega.m $  \\ 
    \hline
  \end{tabular}
  \caption{Characteristics of the constantan-PEEK wires produced by HABIA.\label{tab:HVwire}}
\end{table}

\subsubsection{Internal and external corona}

\begin {figure} 
\centering	
\includegraphics[height=3.0 in]{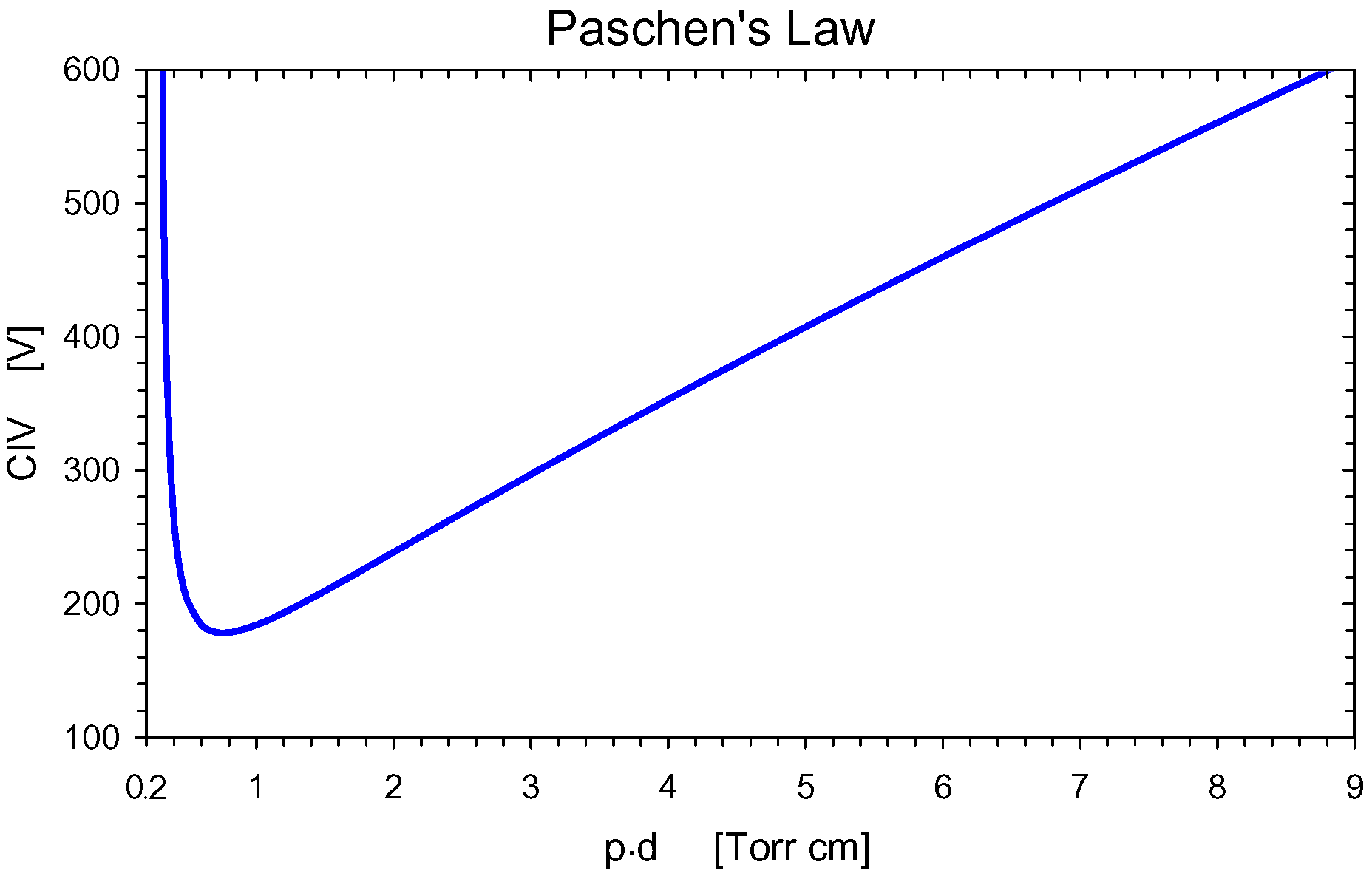}
\caption{Paschen's curve for Ar/Cu.\label{fig:PaschenLaw}}
\end{figure}

Corona is the repetitive charging and discharging of electrical conductors at high voltage through a gas or air.  In the case that one electrode is an insulated HV wire, the corona current and the UV photons created in the electrical avalanche in the gas may slowly damage the insulation and cause catastrophic breakdown. Several good references exist on corona in wiring systems~\cite{perkins72, astm98}.  Corona has the insidious property that it searches out the weakest part of the insulation and eventually drills a hole in it.  The voltage is highest in the gas where the insulation is thinnest around a wire, so that is where the corona will occur. Beginning approximately 100 years ago, extensive studies have been done on the conditions necessary to cause a spark in a gas. It was found experimentally that the sparking voltage is a function of  pressure times distance and followed Paschen's Law, see Fig.~\ref{fig:PaschenLaw}. This curve shows that there is a minimum voltage for corona ignition at any separation of the electrodes. For argon gas  and air this voltage is respectively about 270~V and 330~V \cite{druy40} and is sensitive to both temperature and humidity. 

As described in Ref.~\cite{astm98}, most commercial corona tests are done with alternating current (AC). An AC voltage is applied to the object (considered as a capacitance) under test. The capacitance under test is observed with an oscilloscope through a high-pass filter, designed to filter out the HV AC component of the output signal, while passing the high-frequency corona pulses. In order to avoid internal corona, arising in gas trapped between the conductor and insulation and inside the insulation itself, HV wire manufacturers will try to exclude gas inclusions during the extrusion process. External corona occurs outside the insulation, but both types of corona could, in principle, damage the insulation. HABIA arranged for an outside lab (a swedish national laboratory) to test for internal corona after we noticed, under microscope, an unusually high number of air bubbles in the insulation of the production cable (compared to the thinner prototype wire which had no bubbles).

\begin {figure} 
\centering	
\includegraphics[height=3.0 in]{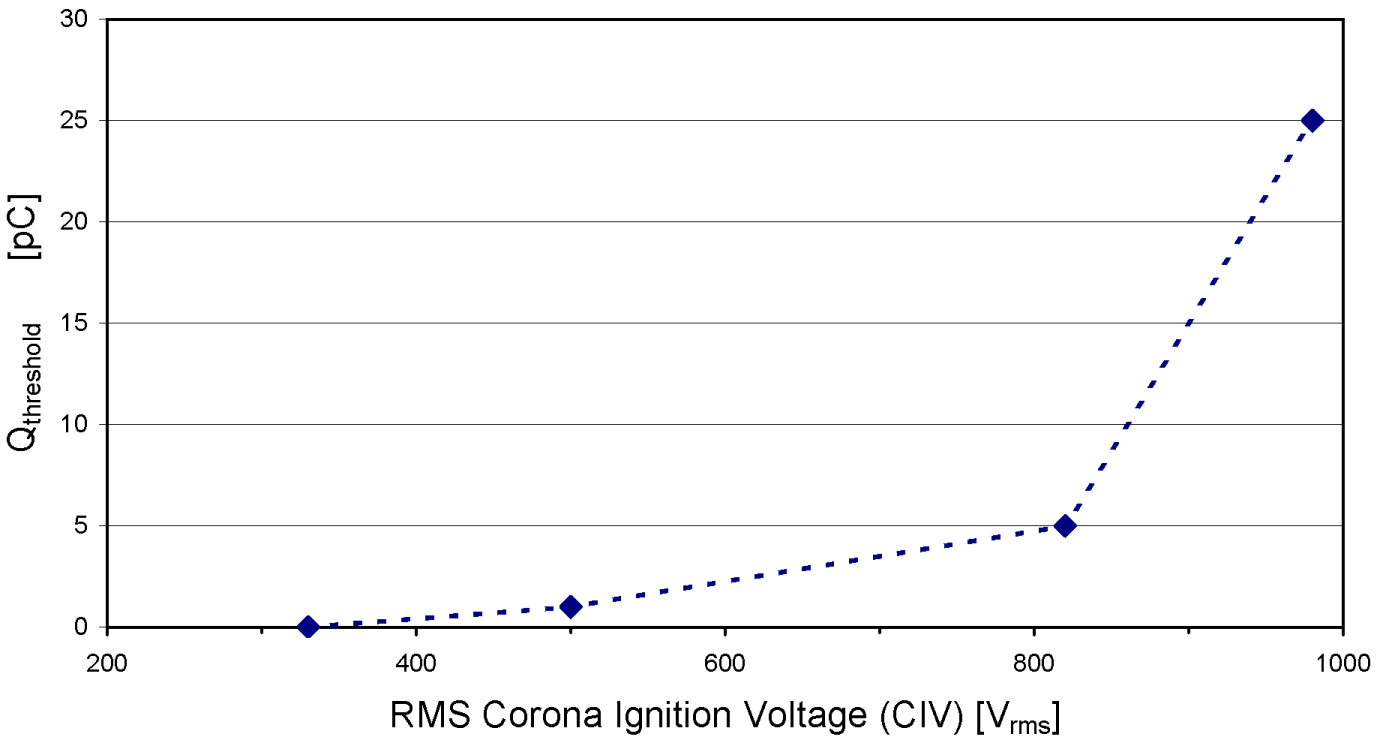}
\caption{Corona charge threshold ($Q_{threshold}$) versus Corona Ignition Voltage (CIV) measured with the HABIA wire wound on a grounded Al cylinder.\label{fig:CIVmeasured}}
\end{figure}

In order to obtain corona ignition voltage (CIV) measurements and to compare with those supplied by HABIA, we conducted our own tests using various geometries at NOVACAP\footnote{NOVACAP, {\tt www.novacap.com/}}. We used 12~m of ATLAS production wire wound on an aluminum mandrel in air. Thus, this test was sensitive to both internal and external corona. Fig.~\ref{fig:CIVmeasured} shows the CIV as a function of the charge threshold in pC (pC) for a single corona pulse. This curve is consistent with the Paschen minimum of 330~V DC or  233~V RMS for very small single-pulse charges in air (the bubbles in the wire contain air that could not be removed in the extrusion process.) The curve reaches 1~pC at 500~V and increases very rapidly after 800~V. The capacitor industry's standard threshold is 25~pC for the single pulse charge threshold. Corona charge pulses smaller than this value are not considered dangerous because they are observed not to damage capacitors.

\begin{table}[htbp]
  \centering
  \begin{tabular}{|l|c|c|c|c|}
    \hline
    Insulation & Test location & Medium & CIV RMS (V) \\ 
    \hline
    Bubble PEEK & Sweden & water & 868, 898\\ 
    Bubble PEEK & NOVACAP & water & 740, 770\\ 
    Bubble PEEK & NOVACAP & mandrel in air & 820\\ 
    Bubble PEEK & NOVACAP & mandrel in epoxy & 610\\ 
    Bubble PEEK & NOVACAP & air (10 cm from ground) & no corona to 3.5 kV\\ 
    \hline
  \end{tabular}
  \caption{CIV for ATLAS production wires at $Q_{threshold}$ = 5 pC.
  \label{tab:CIV}}
\end{table}

Table~\ref{tab:CIV} shows the CIV values measured for ATLAS production wire (Bubble PEEK) both by HABIA in Sweden and by us at NOVACAP. For the test results shown, ``water'' means grounded conducting water ($5\%$ salt by weight), ``mandrel'' means grounded aluminum mandrel in air and ``epoxy'' means conducting epoxy. It can be inferred from the data that there would be no internal corona between adjacent HV wires (assumed to touch) in our high voltage FTs if the voltage difference were less than 740~V $\times$ 2 $\times$ $\sqrt2$ = 2090~V. Thus the most likely source of internal corona would come from a channel which is set to 0~V (off or spare), in case all wires have the same polarity. Even if such internal corona does exist, accelerated destructive corona tests (for internal corona) at both NOVACAP and SBU showed that there would be no failure of the insulation.

Assuming an AC corona pulse is as destructive as a DC corona pulse, we accelerated the destruction of our PEEK insulation by a factor \[A = \tau/ \bigtriangleup t \]
where $\bigtriangleup t$  is 1/120 s (since there are 2 corona pulses per AC cycle, one of each sign) and $\tau$ is the DC corona time constant for PEEK. When an unshielded, insulated wire is at a DC high voltage, the surface of the insulator reaches the full voltage due to leakage current through the insulation. Treating the PEEK insulator as a conductor, the capacitance $C$ of the wire and the resistance $R$ between the conductor and the outer surface can be calculated in terms of the bulk resistivity $\rho_{R}$, the dielectric constant $K\varepsilon_0$, the inner and outer radii $r_{i}$ and $r_{o}$ of the insulator, and the length $L$ of the wire. \[ R = \frac{\rho_{R} \ln {r_{o}/r_{i}} }{2 \pi L} \quad \text {and} \quad C = \frac{2 \pi L K \varepsilon_{0}}{\ln {r_{o}/r_{i}}}\]

Therefore the time constant $\tau = RC = \rho_{R} K \varepsilon_{0}$ of the wire is independent of the geometry. This time constant, which is the charging constant of PEEK considered as a conductor, can otherwise be derived on very general grounds from the equation of continuity and Maxwell's equation, respectively \[\stackrel{\rightarrow}{\nabla}\cdot\stackrel{\rightarrow}{J}  =  - \frac{\partial \rho}{\partial t} 
\quad \text {and} \quad 
  \stackrel{\rightarrow}{\nabla}\cdot\stackrel{\rightarrow}{E}  =  \frac{\rho}{K\varepsilon_{0}}.\] 
Using $\stackrel{\rightarrow}{J} = \stackrel{\rightarrow}{E}/ \rho_{R}$ one finds that the charge density decays exponentially as $\rho = \rho_{0}e^{-t/\tau}$, where $\tau = \rho_{R} K \varepsilon_{0}$. Hence, using $K = 2.27$ and the minimum value for the resistivity of the wire $\rho_{R} = 1.72\times 10^{13}$ $\rm \Omega. m$ (values provided by HABIA) we obtain $\tau$ =  414~s and $A \geq 50,000$.

In ATLAS, we separate the wire bundle from the surrounding FT port structure (maintaining a separation of at least 5~mm except at the WFT itself, where the outer wires are within 1~mm of the steel housing). Therefore, the wire-to-wire corona problem is dominant. We ground the outside of the wire bundle inside the LAr, thus draining the surface charge from the PEEK where there is essentially no internal corona (because of the temperature) and no external corona in the liquid. The NOVACAP tests including the test with a coil wound on a mandrel in air, indicate that internal corona and external corona behave in a similar manner.  But these tests are not definitive because they were performed in air rather than gaseous argon.

The NOVACAP tests conducted at 1.77~kV (2.5~kV$/\sqrt2$) for 68 hrs (384~years DC equivalent) showed no damage to the coil. It is interesting to note that the corona actually stopped after about 5 hrs, phenomenon confirmed by NOVACAP in later tests. During the SBU tests, a coil in conducting water was undamaged after 20.2~hrs at 1.77~kVAC (115 years DC equivalent in ATLAS). In addition, a coil wound on a grounded aluminum mandrel survived undamaged in air for 84.3~hrs at 1.77~kVAC (481 years DC equivalent) and in gaseous argon for 56~hrs (319 years DC equivalent). This latter test shows that the FT wires will not be damaged by either internal or external corona over the lifetime of the ATLAS experiment.

\subsection{The Wire Feedthroughs}

Wire FTs (WFTs), containing 210-220 wires per WFT, were made by the Douglas Electrical Components (DEC)\footnote{Douglas Electrical Components (formerly Douglas Engineering Company), {\tt www.douglaselectrical.com}.}. DEC using a proprietary method to space and ``pot'' the wires in a stainless steel cylindrical housing of 1338 $\rm mm^{2}$ cross sectional area. Inside the potting, the PEEK insulation is stripped over 1-2 cm to obtain a vacuum seal by direct adhesion between potting epoxy and the solid constantan HV conductor. Four WFTs are mounted in holes in the FT plate of HVFT port. These non-cryogenic WFTs provide the required wire density and have a rated wire-to-wire breakdown voltage of 5.0 kV DC. At delivery, the wire bundle on the cold side was up to 10~m long (LAr barrel WFTs), and 0.6~m on the warm side. The WFT contained from 105 to 110 wire ``loops'', where each wire is ``looped'' on the cold side of the WFT. 
DEC tested every wire loop with all other wire loops grounded before shipping the WFT. DEC also guaranteed that the WFT has a leak rate below $10^{-9}$cc/s of He at STP before shipment. The WFT has a radiation tolerance of 400~Gy =  $4\times10^{4}$~rad. 

\subsubsection{Cleaning the WFTs and wires}

An extensive and labor intensive cleaning process, which lasted about two months, took place upon reception of the WFTs from DEC. This cleaning process began with a visual and microscopic inspection of the entry and exit areas of each WFT to check for nicks, cuts, sharp bends and debris, the latter of which are removed by a first water bath. Then, each WFT was immersed in an ultrasonic bath of water and an all-purpose detergent\footnote{Product no.~0037, {\tt www.zepmfg.com}.} at $50^{o}\rm C$ for  at least 2~hrs. All wire surfaces were cleaned, using a minimum of 10 brush stokes, in a third bath made of a mixture of water, detergent and ethyl-alcohol. In the following stage, each WFT was soaked for at least 2 hrs in a mixture containing water and $33\%$ ethyl-alcohol. Finally, all wires were individually cleaned by hand using soft, highly absorbent and non-abrasive wipes (an average of 4 wipes per wire), and 200 proof ethyl-alcohol. This was done until no visible residue is left on the wipes. WFTs were handled exclusively inside a clean room.

\subsubsection{The wire bundles, connectors, and the HV backplane}

Wiring diagrams were made showing the detailed routing path and the corresponding wirelength calculation (including 0.2~m extra for safety) for all HV wires as function of HVFT position and calorimeter destination patch panel. These length calculations were extensively checked. Bundle lengths vary between 1.7~m and 8.3~m. 
Before making bundles, all identified bad wires were removed. The wires on the warm side were cut to a length of 18~cm and female REDEL\footnote{LEMO/REDEL, contacts type no.~ERA.05.403.ZLL1 for 50-contact rectangular chassis connector type no.~SLA.H51.LLZG} contact pins were crimped on the conductors. The crimp connections are protected with a HV lacquer\footnote{HumiSeal, {\tt www.humiseal.com}, lacquer type 1B73.} and a shrink 
tube\footnote{Kynar; polyvinylidene-difluoride, {\tt cableorganizer.com/heat-shrink/}}. 
Then, following the wiring plan and starting from the cold side of the WFT, separate wire bundles of 4-8 wires were formed over the full wire length on a 10~m long bundling platform (netted surface to allow filtered air to pass through from above) and identified with label ties.  
In bundles that carry negative HV (for purity monitors and for the End Calorimeter Presampler), grounded spare (``shield'') wires are used to separate wires carrying negative voltage from the majority of wires carrying positive HV inside the WFT and the Argon gas column. 

A typical WFT contains about 50 bundles, of four wires each. Because of its high radiation resistance of up to 100 Mrad, only Tefzel cable ties\footnote{Tefzel; ETFE - ethylene-tetrafluoroethylene, \\
{\tt www2.dupont.com/Teflon\_Industrial/en\_US/assets/downloads/h96518.pdf}.} were used for all bundling operations. After bundle formation, the bundles were cut to size ($\pm 1$~cm) and connectorized with the cold connectors\footnote{Single-contact custom design in PEEK material for LAL, 91405 Orsay, France, by ATI Electronique (Alliance Technique Industrielle), {\tt www.ati-electronique.fr}, used for most calorimeters; and 7-contact (4 used) glass-filled polyester connector by AMP/Tyco Electronics, USA, {\tt www.tycoelectronics.com}, part no.~1-87175-5, used for the hadronic End Calorimeters.}. Grounding a finished bundle on the cold side, its corresponding wire contacts on the warm side were located using a powered LED pen, and inserted into the LEMO/REDEL chassis connector block. 

\begin {figure} 
\centering
\includegraphics[height=3.0 in]{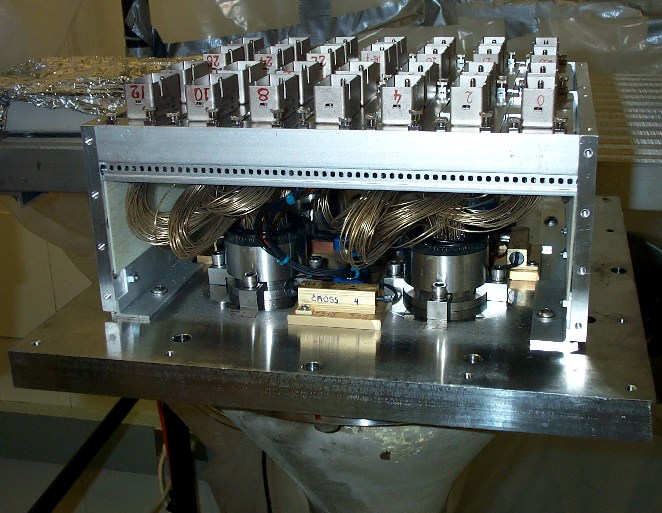}
\caption{View of a FT plate with WFTs, bolted on the HVFT reducer flange. In the foreground one of four heater resistors is seen. The HV backplane with four rows of seven REDEL multi-contact HV connectors is mounted over the FT plate.\label{fig:HVFTplate}}
\end{figure}

The wire bundles were carefully guided through the appropriate FT plate holes and the WFTs were bolted in place. The HV Backplane is mounted on the FT plate. The warm REDEL HV connectors are mounted on the 56-HP-wide (1~HP = 5.08~mm) backplane, constructed using EURO-standard hardware, as shown in Fig.\ref{fig:HVFTplate}. Before leaving the clean room, to prepare the wires for shipment to CERN, the bundles on the cold side were wrapped with protective plastic cling wrap to form a single HV bundle. This bundle was threaded through a PEEK spacer ring (that centers the bundle in the throat of the metal reducer of the HVFT port), looped and attached to a cage supported by the FT plate, and the whole was bagged in plastic for mechanical and dirt protection.

The FT plate, with HV backplane on top and the cold HV bundle below, was then ready for shipping and installation in the HVFT port on the calorimeter cryostat.

\subsubsection{Tests and repair procedures}

As an integral part of the quality control of the HVFTs, each WFT was inspected for nicks, cuts, debris and sharp bends, and a HV test was performed upon reception. For this test, the WFT was immersed in water at ground potential. We applied 4.8~kV to each wire loop, one at a time, while keeping all others at ground. This test discovered shorts to adjacent wires or housing and breaks in the wire insulation of the wires on the warm and cold sides. On average, two shorted (i.e. current $>10$~nA) wire loops per WFT were found this way. These loops were cut and the faulty wires were identified and labeled on both cold and warm side of the WFT. During bundling, the faulty wires were removed. 

\begin {figure} 
\centering	
\includegraphics[height=2.5 in]{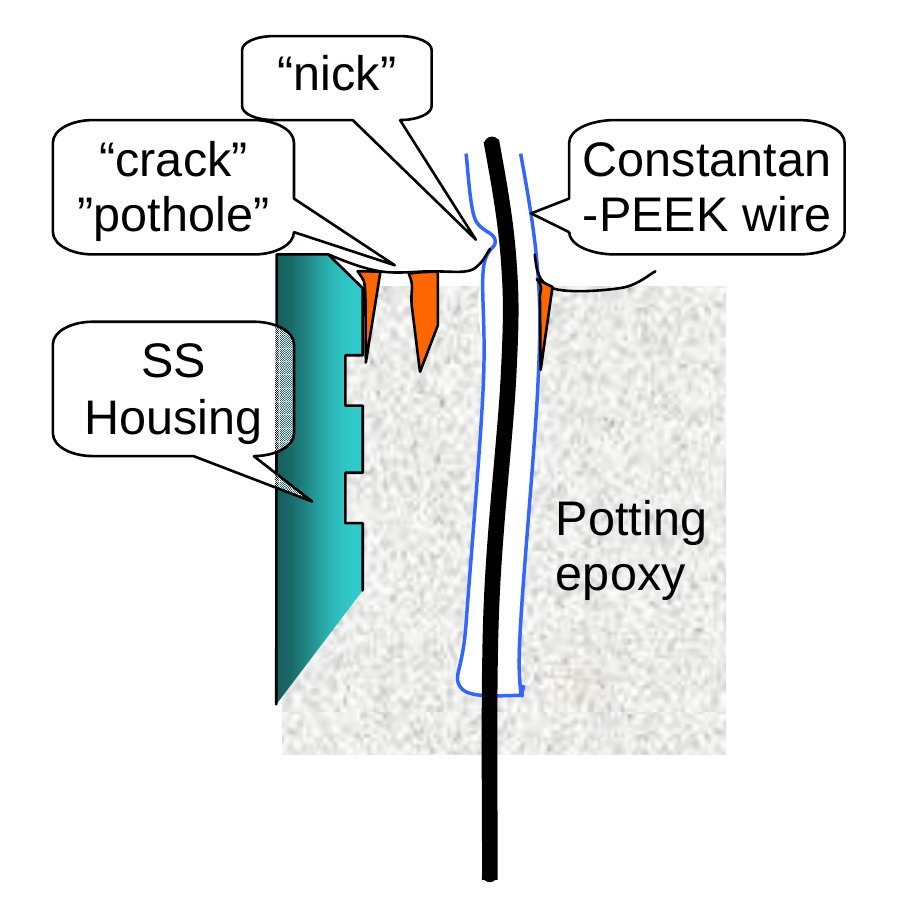}
\hfil
\includegraphics[height=2.5 in]{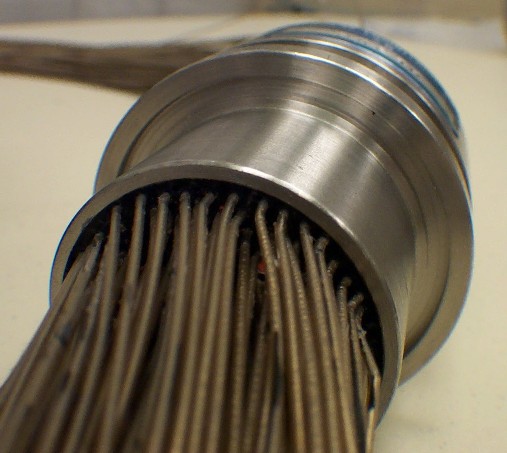}
\caption{An illustration of Wire Feedthrough faults (left). Right: wires exiting a WFT on the cold side, showing the high-density 216-wire HV bundle.\label{fig:WFTdetail}}
\end{figure}

A second, more sensitive and detailed HV test was performed on the WFT at the final stage of production, after wires were cleaned, bundled, and connectorized on both ends. For this test the cryostat side of the wire bundle was enveloped in a plexiglass tube filled with Argon gas and the filter side was bagged in air of 60\% relative humidity. During the final test of the first WFTs, a large number of breakdowns ($\sim 10$ wires per WFT or $\sim 5\%$ of total) were found. A closer inspection of the WFTs showed that failing wires are mostly located around the circumference of a given WFT bundle and near the rim of the stainless steel housing. This inspection also showed cracks in the potting epoxy near the metal housing rim and around the perimeter of some wires, with damage to the PEEK insulation, see Fig.~\ref{fig:WFTdetail}. The failures were most likely caused by frequent handling of the WFTs and wire bundles during manufacture at DEC and later cleaning and bundling operations at SBU. Details of WFT construction (wires exiting close to the metal housing, poor epoxy-PEEK adhesion, excessive degassing, etc.) may have contributed as well.

A repair procedure was put in place because the number of failures was too large to be accommodated by using spare wires. The solution was to re-pot both surfaces of the WFTs in-house with a non-conducting epoxy under vacuum (0.05 atm). Extensive tests of six candidate epoxies led to the choice of the Tra-Bond-2115\footnote{TRA-CON, {\tt www.tra-con.com/products/tpb.asp?product=2115}.}. This expoxy has a low viscosity (250 cps at $25^{0}\rm C$) and a good dielectric strength (6.0 kV). In tests it demonstrated very good adhesion to the original WFT potting epoxy as well as excellent adhesion to the PEEK insulation of the wires. Tra-Bond-2115 was also tested for use in the ATLAS TRT and found to have good mechanical stability and low pollution potential to LAr. Subsequent detailed HV tests of the WFTs at 3.0 kV showed very few remaining failures: about 2 wires per WFT had $>2$~nA (the sensitivity limit of our HV powersupply\footnote{T.\,Droege, Fermi National Accelerator Laboratory, FERMILAB-TM-0527-A, 1975.}) to ground. Remaining failures were all internal to the WFT, and such wires were replaced by spare wires in the WFT.      

\section{HV RC filters}

The HV RC filters are designed for the purpose of reducing noise from the HV power supplies and EM pick-up to below 1~$\rm mV_{pp}$ (peak-to-peak) at frequencies much below the bunch-turn frequency (100~${\rm \mu s})^{-1}$, and to below 1 $\rm \mu V_{pp}$ at signal frequency (25-100~$\rm {ns})^{-1}$. 
Filters have to also smooth out current draw fluctuations due to large showers (on the 100 ns time scale) and bunch-train patterns (100~ns to 100~${\rm\mu s}$).
Cross talk between HV leads to sub detector elements is not considered a problem. First, fluctuations in current drawn by an element are reduced by filtering at the patch panels and on the modules: the shielding of cable harnesses serves as a distributed capacitance, and several resistors of 1~$\rm M\Omega$ are in series to the calorimeter pads. Second, interline capacitance is about 2 pF/m for wires spaced by 4~mm. The cross talk percentage then roughly equals 30~pF divided by the 27~nF of the filter capacitance to ground, yielding a factor 1000 reduction.	

Because the HV bundles from groups of calorimeter cells are redistributed over different HV power supplies (HVPS) such as to provide operational redundancy (e.g. each half of a LAr gap is powered by a different HVPS), the HV backplane provides a patch-panel function as well as a modular connection to HV filters. Moreover, spare channels needed to replace wires broken during calorimeter installation or to provide separate HV powering to ``weak'' calorimeter cells discovered during calorimeter module construction and testing, are switched in readily by moving contact pins on the backplane. For further compatibility, all warm HV connectors in the LAr system are of the same type LEMO/REDEL $50\times3$~kV (32 channels used per connector).

\subsection{The HV filter module}

\begin {figure} 
\centering
\includegraphics[height=2.2 in]{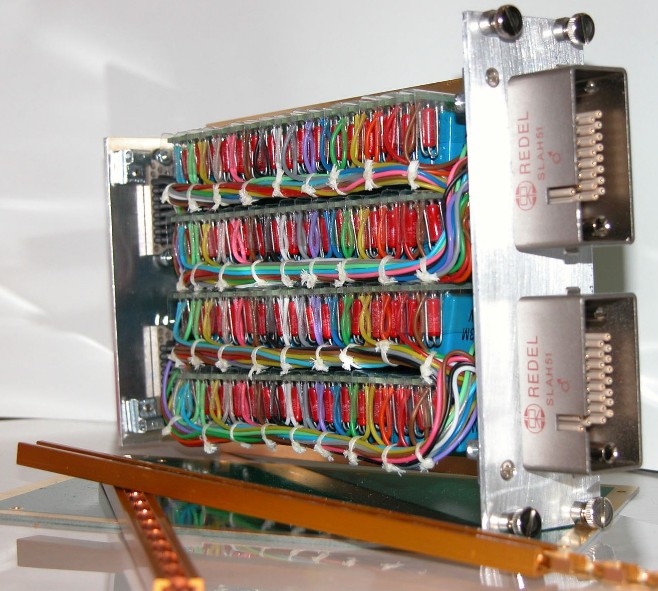}
\hfil
\includegraphics[height=2.2 in]{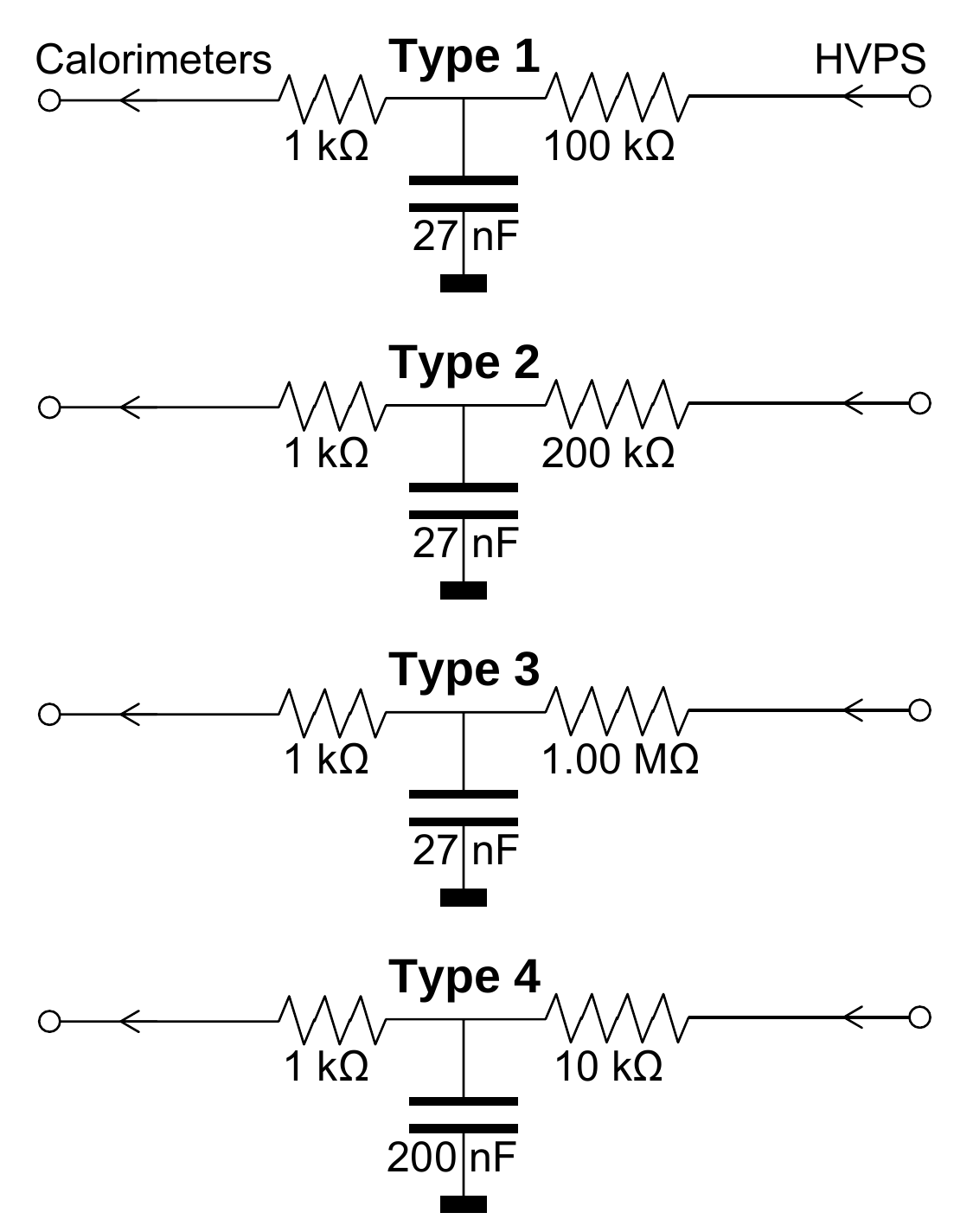}
\hfil
\includegraphics[height=2.2 in]{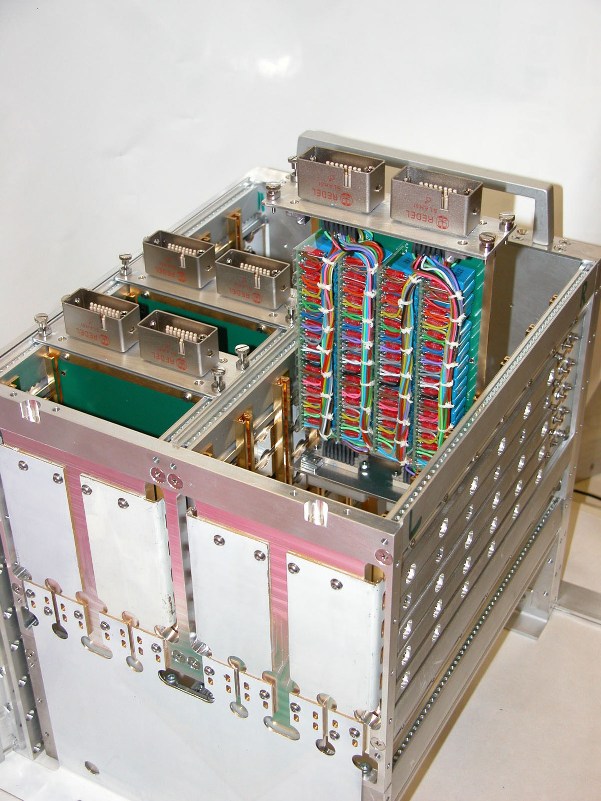}
\caption{Left: a filter module showing the four daughter boards with 16 rCR filter circuits each. The grounding cover is removed, and two Ground-R-Guides card guides are shown in the foreground. 
Middle: arrangement and component values of the four types of HV filters: Type 1 for EM calorimeters, Type 2 for Preshowers and the Hadronic Calorimeters, Type 3 for the Forward calorimeters, and Type 4 for the LAr Purity monitors. 
Right: HV filter crate with three filter modules mounted on a HV backplane and without the shielding side panels in place. The grounding strips (left side of photograph at the ``bottom'' of the crate) and the ground planes leading from the strips to the FT plate (right-front side of photograph and at the side of the crate) are clearly visible. \label{fig:FilterModule}}
\end{figure}	

The HV lines for each FT port are distributed over 12 to 14 HV filter modules of 64 HV filters each. The module is made of a motherboard on which four daughter boards reside mounted at right angles, each with 16 filter sections. The module dimensions follow the EURO standard: 3U high (5.25~in.) $\times$ 8HP wide $\times$ 165~mm deep, and use EURO hardware. Besides serving as mechanical supports for the filter cards, the motherboards also serve as a low-inductance ground plane via grounding card guides\footnote{Ground-R-Guide, Unitrack Industries Inc., {\tt www.unitrackind.com}.} in the filter crate.

As shown Fig.~\ref{fig:FilterModule}, the filters are mounted in a very tight configuration. The filter circuits are separated from each other by thin sheets of Mylar\footnote{Mylar by DuPont, {\tt www2.dupont.com/Products/en\_RU/Mylar\_en.html}.} mounted in slots between neighboring circuits. For the required flexibility, the HV wire used was multi-strand tin-plated copper with polyethylene insulation (the same wire as in the multi-conductor HV cables for ATLAS). Special care was taken in the soldering (avoiding all sharp points) and intensive ultrasonic cleaning was performed to minimize surface conduction. Mother boards, card guides and grounding rims are gold-plated, and RF gaskets are used on grounding connections and around frontpanels in order to provide shielding and a low-inductance ground path for the filters. 

Each RC filter is made of a 27~nF, 5~kV multilayer TiO$_2$ class X7R ceramic encapsulated radial-lead capacitor and two 1\% 0.5~W carbon resistors (a small ``blocking'' resistor and the main filter resistor) whose values are detector specific. The resistors were tested to remain stable under over-current conditions (0.9 W for 1 week), and after repetitive charge pulses (0.2 Hz, 2~kV, 10~$\rm\mu$F for 24 hrs). 
The ATLAS LAr HV filters require nearly 5000 HV capacitors. The reliability of the filter capacitors is a concern, because access for repair/replacement will only be possible once per year during a shutdown. If a capacitor shorts to ground, the corresponding HV channel becomes unusable. Therefore, in order to obtain superior reliability, we use the same, reliability-tested, type of capacitor everywhere. The filter capacitors were purchased from NOVACAP\footnote{NOVACAP, part no.~6560B273M502LEH, {\tt www.novacap.com}.}. NOVACAP tested all capacitors at 6~kV and 125~$^o$C for 100~hrs, in accordance with the MIL-PRF-49467A protocol. From the results of these tests and derating with voltage, we extrapolate a failure rate of less than 1 failure per 10 years of operation for the entire calorimeter. 

The radiation environment of the filter capacitors in ATLAS is at the level of a total maximum integrated dose of 50~kRad and 1$\times10^{13}$ $\rm neutrons/cm^{2}$. This includes a safety factor of 2 to 5. A prototype filter module (64 capacitors) was tested in CERI (France). The module, positioned 80 cm from the source, received a total dose of 5 $\times10^{12}$ neutrons/cm$^{2}$ of 6~MeV while at 5.0~kV. No breakdown was observed, and the total leakage current ($80\pm30$ nA before radiation) was $50\pm20$ nA after radiation. We have not found any data or experiences of multilayered capacitor failures caused by radiation in the MIL/AEI literature. 

The completed filterboards were extensively tested by applying 3.5~kV and ground to alternating filter sections. All individual filter sections of the board, as well as the combined HV input and ground lines were monitored by a capacitively coupled oscilloscope and by nA current meters. The measurements were performed inside a Faraday cage for greatest sensitivity. It was observed that for good boards the leakage current per daughter board decreases to below 20~nA and that the number of charge pulses per daughter board rapidly decreases within a few minutes. Filter boards were accepted if they exhibited a discharge rate below 15~pulses/3~mins, where a charge pulse is defined to have 1~pC or more; at this level, the long-term polarization of the X7R dielectric dominates and is indistinguishable from discharges. 1~kV capacitors for type 4 filters were tested at 800~V. 

Filter modules serve HV channels of similar voltage and polarity.
Four different types of filter modules, see Fig.~\ref{fig:FilterModule}, are used to accommodate the different subsystems in the LAR. Module types 1, 2 and 3 have 27~nF 5~kV capacitors. The choice of filter resistor value depends on the expected current draw in the part of the calorimeter supplied by the HV channel, and is a compromise between best noise rejection and minimum voltage droop across the resistor. With R = 200~k$\Omega$ the rejection is $6 \times 10^{-7}$ for 50~MHz noise. 
Two pairs of LEMO/REDEL connectors are used on the front panel (female) and the back panel (male) of each filter module and have the exact same pin-out as the female REDEL connectors on the HV backplane. As on the backplane, 32 out of 51 pins in each connector are used. On the frontpanel, the unused central row of contact holes in the connector is used for interlock purposes to ensure proper routing to the HVPSs. This arrangement permits easy exchange of filter modules.

In addition to the filter modules, each feedthrough crate also contains a heater module. The heater module filters the currents to four resitive heaters and the signals from two thermocouples which are mounted on the FT plate.

\subsection{The HV filter crates}

Each filter crate is constructed using EURO-standard hardware and forms a 56-HP wide ($\sim$28~cm) $\times$ $\sim$27~cm anodized aluminum structure whose height extends to about 30~cm above the FT flange. It houses two rows of seven  filter modules, which can be removed and replaced easily. It is mounted on top of the FT flange above the HV backplane, see Fig.\ref{fig:FilterModule}. Four aluminum side panels of the crate, together with the RF gaskets on the filter module front panels, form a Faraday cage around the HV lines and filters, while providing good access to the HV backplane (by removing the top and/or bottom panels of the crate).
Low inductance grounding contact is provided internally by wide gold-plated Cu strips between the grounding card guides and the FT plate (via a Cu-Be gasket). Shielding is provided by a Faraday cage formed by outer Aluminum panels and the filter module front panels with RF gaskets. 

\section{Installation and Commissioning}

Installation of the HVFT port mechanics started early January 2002 on the LAr barrel cryostat. By June of the same year the HVFT ports for the end caps were at CERN and ready for installation. Each cryostat was vacuum and pressure tested with all HVFT ports, with blank FT plates, in place. Thorough cleaning of the cryostats was performed before calorimeter installation started. Visible dirt was removed and all surfaces wiped down with ethyl alcohol or other appropriate solvents. The remaining part of the installation took place under a protective tent.

The FT plates, with the wire bundle attached to a harness cage and a strain relief near the FT plate were shipped and installed separately at the time of calorimeter insertion at CERN. Scaffolding and a jib crane were used to lift the FT plate and wire bundle to the top of the cryostat. The wire bundles were gradually released from their supporting cage and fed through the FT port, while the protective cling wrap was being removed. Near the end of insertion, the strain of the cable weight was transferred to attachments points inside the cryostat and the strain relief on the FT plate was removed.  At this point the FT plate, the O-ring seal and RF gasket were inspected (and cleaned if needed) and the FT plate was fully lowered and bolted onto the HVFT port flange. Following the wire routing plan, the individual wire bundles were run around the outer perimeter of the cold cryostat vessel until they reach the azimuth of their destination patch panel, fixing the bundles every 20~cm with Tefzel cable hooks and ties. From the destination azimuth the bundles run radially inward to the module patch panels. The cold HV connectors (ATI) remained bagged and temporarily fixed awaiting the last HV test before connecting.

For this test, each HV contact on the HV backplane was put at 3~kV with all other contacts grounded. The feedthrough pipe was bagged from below and filled with argon gas.  A final prototype ISEG 32-channel HV power supply controlled by a laptop PC was used to perform the tests and read currents (10~nA resolution). Spare cables are available to fix HV problems while access in the cryostat is possible. During HV wire routing and calorimeter connection, two HV cables were accidentally damaged and were successfully replaced.

The detailed HV connectivity was tested several times during calorimeter installation, and no routing mistakes were found. Weak calorimeter channels found during HV tests during installation were connected to spare wires that were routed to strategical positions around the circumference; such channels can be supplied with lower HV if so needed.

Filter crates and filter modules were installed after closing and before the first cooldown of the cryostats. Some failures have been observed and repaired since then. Although all REDEL pin contacts were mechanically tested after insertion, four times a HV contact in a HV back plane or filter module connector was found to have been pushed back. Twice a HV wire break was found at the wire crimp under the backplane, most likely a result of an accidental disturbance before or during the installation of the filter crate. All these failures were successfully repaired. No failures of filter components have been observed at the time of writing, after about 1 year of HV operations.

\section{Conclusions}

High Voltage FTs have been constructed to provide 5040 individual high voltage lines for the ATLAS LAr calorimeters. The non-accessible part of the system is designed to operate without failures for at least 20 years. The failure rate of accessible components (HV filter components) is expected to be less than 1 failure in 10 years.

\section{Acknowledgements}
We thank our staff at both Stony Brook and Brookhaven, and our ATLAS collaborators in the LAr group for their help with this effort. We acknowledge the efforts of the late Dr. F. Lobkowicz, who started the design of the ATLAS HVFTs. We thank Dr. H. Braun for helpfull discussions and assistance; and Drs. M. Aleksa, L. Hervas, P. Fassnacht, and P. Pailler for their insights and support. This work was partially supported by DOE grant DEFG0292ER40697 and NSF grant PHY0652607.

\end{document}